\documentclass[prd,aps,amsfonts,superscriptaddress,nofootinbib,notitlepage,longbibliography]{revtex4-1}

\usepackage{graphicx}
\usepackage{tabularx}
\usepackage{physics}
\usepackage{amssymb}
\usepackage{amsmath}
\usepackage{bm}
\usepackage{stmaryrd}
\usepackage[driverfallback=dvipdfm]{hyperref}
\usepackage[dvipsnames]{xcolor}
\hypersetup{colorlinks=true,breaklinks,urlcolor=blue,linkcolor=blue,citecolor=red}

\newcommand{\comment}[1]{}

\makeatletter
\renewcommand{\p@subsection}{}
\renewcommand{\p@subsubsection}{}
\makeatother

\usepackage{amsthm}

\numberwithin{thm}{section}

\begin{document}

\title{Quantum error correction in a time-dependent transverse field Ising model}

\author{Yifan Hong}
\email{yifan.hong@colorado.edu}
\affiliation{Department of Physics, University of Colorado, Boulder CO 80309, USA}
\affiliation{Center for Theory of Quantum Matter, University of Colorado, Boulder CO 80309, USA}

\author{Jeremy T. Young}
\affiliation{Department of Physics, University of Colorado, Boulder CO 80309, USA}
\affiliation{Center for Theory of Quantum Matter, University of Colorado, Boulder CO 80309, USA}

\affiliation{JILA, University of Colorado and National Institute of Standards and Technology, Boulder, CO 80309, USA}

\author{Adam M. Kaufman}
\affiliation{Department of Physics, University of Colorado, Boulder CO 80309, USA}

\affiliation{JILA, University of Colorado and National Institute of Standards and Technology, Boulder, CO 80309, USA}

\author{Andrew Lucas}
\email{andrew.j.lucas@colorado.edu}
\affiliation{Department of Physics, University of Colorado, Boulder CO 80309, USA}
\affiliation{Center for Theory of Quantum Matter, University of Colorado, Boulder CO 80309, USA}

\date{\today}

\begin{abstract}
We describe a simple quantum error correcting code built out of a time-dependent transverse field Ising model.  The code is similar to a repetition code, but has two advantages: an $N$-qubit code can be implemented with a finite-depth spatially local unitary circuit, and it can subsequently protect against both $X$ and $Z$ errors if $N\ge 10$ is even. We propose an implementation of this code with 10 ultracold Rydberg atoms in optical tweezers, along with further generalizations of the code.
\end{abstract}

\maketitle

\tableofcontents

\section{Introduction}

Finding fast and efficient protocols for quantum error correction that can be implemented in present-day quantum platforms (superconducting qubits \cite{PhysRevLett.89.117901, PhysRevLett.79.2328, Krantz_2019, Brooks_2013, Gu_2017}, trapped ions \cite{PhysRevLett.74.4091, Bruzewicz_2019, kihwan_2010, Britton_2012, Barreiro_2011}, Rydberg atoms \cite{Saffman2010, Saffman2016, Browaeys2020, Wu2021, Morgado2021}, cavity quantum electrodynamics \cite{PhysRevLett.104.073602, science.aar3102, PhysRevLett.125.060402, RevModPhys.85.553}, photons \cite{Wang_2018, Wang_2019}, silicon \cite{Kane1998ASN, He_2019}) is a problem of widespread interest.  Perhaps the most intuitive model of quantum error correction is the quantum repetition code \cite{peres_1985}, which can correct effectively against a single type of error (which we take to be $Z$).  Given an initial state consisting of $N$ qubits \begin{align}\label{eq:psi0}
    \ket{\Psi_0} = \big( \alpha\ket{0} + \beta\ket{1}\big) \otimes \ket{00\dots 0} \; ,
\end{align}
where $|\alpha |^2 + |\beta |^2 = 1$ for normalization and $\ket{0},\ket{1}$ are local spin-$\uparrow$, spin-$\downarrow$ states in the Pauli-$Z$ basis respectively, one can find a unitary $U_{\mathrm{QRC}}$ such that \begin{equation}\label{eq:QRC}
   \ket{\Psi_{\mathrm{QRC}}} =  U_{\mathrm{QRC}} \ket{\Psi_0} = \frac{\alpha}{\sqrt{2^{N-1}}}\sum_{\text{even parity } \mathbf{s}} \ket{\mathbf{s}} + \frac{\beta}{\sqrt{2^{N-1}}}\sum_{\text{odd parity } \mathbf{s}} \ket{\mathbf{s}}
\end{equation}
where a bit string $\mathbf{s}$ is even/odd parity if it has an even/odd number of 1s.  We can think of $\ket{\Psi_{\mathrm{QRC}}} $ as a \emph{parity-check state}: the parity of the strings determines whether the coefficient is $\alpha$ vs. $\beta$.  This parity-check nature makes it easy to correct against $Z$ measurements.  For example, if measuring $Z$ on the last qubit, if the outcome is 0, then we simply retain the information in the other $N-1$ qubits; if the outcome is 1, the information is still stored, but we need to apply an $X$ gate at the end to recover the original qubit.

A key shortcoming of this model is its inability to correct against even a single $X$ measurement, which collapses the entire wave function.  Of course, more sophisticated codes \cite{Gottesman:1997zz} are known which can protect against both a $Z$ and $X$ error; simplest conceptually among them the Shor 9-qubit code \cite{shor_1995}.  More practical possibilities include the surface code \cite{Kitaev:1997wr,Bravyi:1998sy,Dennis:2001nw,fowler2,fowler}, which is more amenable to physical implementation (and more fault tolerant); at least 9 data qubits are needed to protect one logical qubit in the surface code \cite{fowler}.

In this paper, we present another simple alternative to the quantum repetition code, which solves two shortcomings of the repetition code, while maintaining most of its conceptual simplicity.    Our code is generated by a one-dimensional, spatially local, time-dependent transverse field Ising model (TFIM).  While this model has a celebrated history in quantum information theory due to its connection with proposed Majorana-based quantum computation \cite{Kitaev:2000nmw,Bravyi:2010de,dassarma,Mourik:2012xn,Nadj_Perge_2014}, here we will point out a rather different way that the TFIM can be used to encode a qubit robustly.  Like the repetition code, our code is inspired by the use of parity-check states to effectively correct against $Z$ measurement/errors.  Indeed, connections between (random) transverse field Ising model dynamics and quantum error correction in the repetition code have been emphasized in \cite{lang,hsieh,Li:2021bph}.  Unlike the repetition code, which relies on the preparation of a GHZ state, our parity-checked state can be prepared in constant time under unitary dynamics, and it leads to a code which can correct against both $Z$ and $X$ errors.  The ability of our code to achieve such error-correcting parity-checked states after finite time unitary dynamics can be understood through a connection with symmetry-protected topological (SPT) phases \cite{2004.07243,Verresen:2021wdv,Tantivasadakarn:2021vel}, although this code appears simpler than many others inspired by condensed matter physics. 

The TFIM code we present is naturally realized using the recent progress made in the control and manipulation of quantum systems. In particular, optical tweezer arrays of Rydberg atoms have proven  to be a highly tunable system for quantum applications due to the ability to control the atoms individually \cite{Bernien2017, Lienhard2018, Keesling2019, Norcia2019, Browaeys2020, Scholl2021, Schine2021}. Furthermore, while controlling the initial spatial configuration of the atoms is already a powerful tool, it is now also possible to move the atoms while preserving qubit coherence \cite{Bluvstein2021}. This high degree of control, both in space and time, position optical tweezer arrays as an excellent platform for realizing the TFIM code in near-term experiments.

The rest of the paper is organized as follows: we will introduce the TFIM code in Section \ref{sec2}.  In Section \ref{sec3} we describe conventional syndrome-based quantum error correction, and show how the TFIM code both recovers the more conventional phenomenology of the repetition code in the presence of Z errors (in our basis) and can also go beyond it by correcting X errors. We present numerical evidence in Section \ref{sec5} that the TFIM code can straightforwardly be used to generate higher depth codes.  The feasibility of implementing the TFIM code in ultracold atom experiment is described in Section \ref{sec6}.

\section{Unitary encoding}\label{sec2}

We begin with a 1D spin-chain of uncoupled qubits in a product state with our information ($\alpha,\beta$) encoded in $\ket{\psi} = \alpha\ket{0} + \beta\ket{1}$ on the first site, as in (\ref{eq:psi0}).  We wish to encode this logical qubit amongst $N$ physical qubits in order to protect the information against unwanted projective measurements.

\subsection{Transverse-field Ising model}

\begin{figure}[t]
\centering
\includegraphics[width=.38\textwidth]{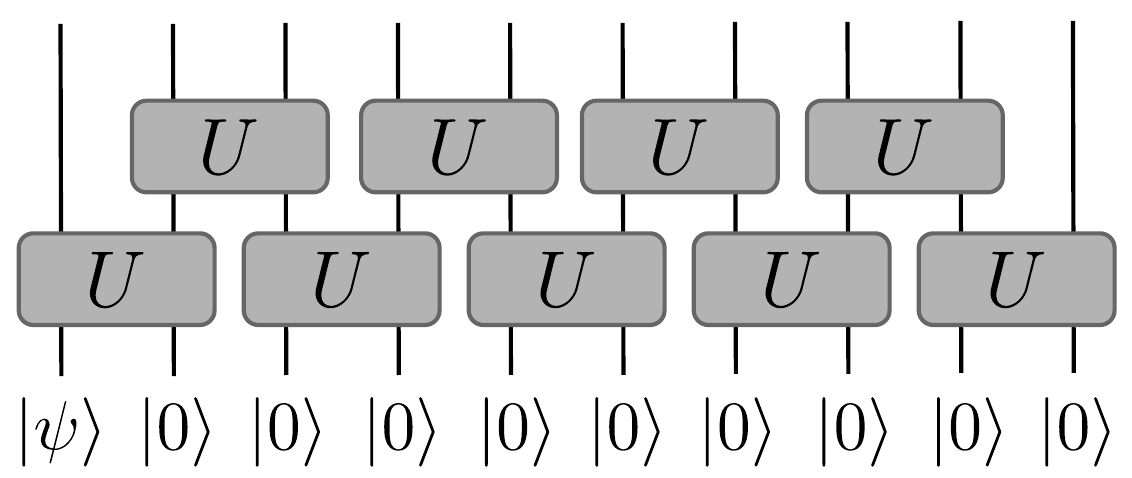}
\caption{Illustration of the nearest-neighbor encoding procedure for $N=10$ sites.}
\label{fig:encoder}
\end{figure}

The encoding procedure consists of parallelized nearest-neighbor unitary gates on alternating even and odd bonds illustrated in Fig. \ref{fig:encoder}. Each unitary gate beginning on a spatial site $i=1,2,\ldots, N$ is given by
\begin{align}\label{eq:gate}
    U_i = \frac{1}{\sqrt{2}}\left( Z_i + X_iX_{i+1}\right) \; .
\end{align}
The first term $Z_i$ can be thought of as a transverse-field interaction in addition to the ferromagnetic coupling in the second term $X_iX_{i+1}$.  Since this gate is involutory $(U_i^2=1)$, it can be generated by itself up to an overall phase:
\begin{align}
    U_i = \mathrm{i} \mathrm{e}^{-\mathrm{i}\frac{\pi}{2}U_i} \; .
\end{align}
Thus, we can realize this encoding procedure with a time-dependent transverse-field Ising model (TFIM) Hamiltonian
\begin{align}
    H_\mathrm{TFIM}(t) = \sum_{\langle ij \rangle} J_{ij}(t) X_iX_j + \sum_{i=1}^N h_i(t) Z_i \; ,
\end{align}
where $\langle ij \rangle$ denote spatially adjacent sites $i$ and $j$ in the 1d lattice (i.e. $|i-j|=1$). For our code, we desire equal coupling strengths $h_i(t)=J$, while $J_{ij}(t)$ will alternate between values $J$ and $0$, according to the circuit sketched in Fig. \ref{fig:encoder}.  The runtime of each layer of the circuit is $\mathrm{\Delta}t = \pi/2\sqrt{2}J$.

It is likely not accidental that the strength of the transverse field and the Ising term are equal during application of a gate.  Indeed, if we simply turned on all couplings for all time, this would be the TFIM tuned to its quantum critical point, which has been known to exhibit useful error correcting properties \cite{Pastawski:2016ggn} due to its relationship with conformal field theory (CFT).  However, the actual decoding process (i.e. codebook) generated by this CFT would not be practical to implement; the code we present, in contrast, will be implementable.

All of our unitary gates commute with the global $Z$-parity operator
\begin{align}\label{eq:Zbar}
    \bar{Z} = \prod_{i=1}^N Z_i
\end{align}
so we can think of $\bar{Z}$ as a global symmetry operator of the entire circuit. After the encoding procedure, the quantum state will take the parity-check form
\begin{align}\label{eq:PC state}
    \ket{\Psi_\mathrm{PC}} = \frac{\alpha}{\sqrt{2^{N-1}}} \sum_{\text{even strings }\mathbf{s}} \eta_{\mathbf{s}} \ket{\vb{s}} + \frac{\beta}{\sqrt{2^{N-1}}} \sum_{\text{odd strings }\mathbf{s}} \zeta_{\mathbf{s}} \ket{\vb{s}} \; ,
\end{align}
where the even/odd bitstring states are eigenstates of $\bar{Z}$ with eigenvalues $\pm1$, and $\eta_i,\zeta_i = \pm 1$ are phase factors determined from the encoding procedure which are not important (for us) to determine directly: we will instead keep track of signs in the logical qubit operators $X_L$ and $Z_L$ (which allow us to recover an arbitrary qubit).

\subsection{Heisenberg picture}

In subsequent analyses, we will compute expectation values of various Pauli operators with respect to our quantum state. In particular, the expectation value of some local operator with the initial state will in general become that of a nonlocal one with the time-evolved state:
\begin{align}\label{eq:reverse time evol}
    \expval{O_i}{\Psi_0} = \expval{\tilde{U}^\dagger \tilde{U}O_i\tilde{U}^\dagger \tilde{U}}{\Psi_0} = \expval{\tilde{U}O_i\tilde{U}^\dagger}{\Psi(t)} \; ,
\end{align}
where $\tilde{U}$ is the initial encoding procedure. We see that our initial local operator $O_i$ has undergone reverse time evolution under $\tilde{U}$, defined as \begin{equation}
    \tilde{U} = U_2 U_4\cdots U_{N-2} \times  U_1 U_3 \cdots U_{N-1} .
\end{equation} Since our initial state (\ref{eq:psi0}) is a product state, it is a +1 eigenstate of an exponentially large number of local operators called the initial stabilizer. After time evolution, the state will be a +1 eigenstate of the reverse time-evolved operators according to (\ref{eq:reverse time evol}). Thus, measuring these ``check operators" after time evolution is equivalent to measuring the initial stabilizer.

We list the action of $\tilde{U}$ on a few local operators below, which will be of convenience later: 
\begin{subequations}\label{eq:28}\begin{align}
    \tilde{U}X_k\tilde{U}^\dagger &= \begin{cases}
        -Y_{k-1}Y_kZ_{k+1}X_{k+2} & \text{ for odd $1<k<N-2$} \\
        Z_kX_{k+1} & \text{ for even $k<N$}
    \end{cases} \\
    \tilde{U}Y_k\tilde{U}^\dagger &= \begin{cases}
        -Y_{k-1}Z_k & \text{ for odd $k>1$} \\
        Y_{k-2}Z_{k-1}X_kX_{k+1} & \text{ for even $2<k<N$}
    \end{cases} \\
    \tilde{U}Z_k\tilde{U}^\dagger &= \begin{cases}
        X_kZ_{k+1}X_{k+2} & \text{ for odd $k<N-2$} \\
        Y_{k-2}Z_{k-1}Y_k & \text{ for even $2<k$} \; .
    \end{cases}
\end{align}\end{subequations}
We also list the action of an individual 2-site unitary $U_i$ acting on sites $i$ and $i+1$ (recall that $U_i^\dagger = U_i$):
\begin{subequations}\label{eq:29}\begin{align}
    U_iX_iU_i &= Z_iX_{i+1} \\
    U_iY_iU_i &= -Y_i \\
    U_iZ_iU_i &= X_iX_{i+1} \\
    U_iX_{i+1}U_i &= X_{i+1} \\
    U_iY_{i+1}U_i &= Y_iZ_{i+1} \\
    U_iZ_{i+1}U_i &= -Y_iY_{i+1} \; .
\end{align}\end{subequations}
We readily observe that the TFIM code is a Clifford circuit and is thus easy to simulate \cite{gottesman1998,Aaronson_2004}.  (\ref{eq:29}) can also be used to readily determine the edge cases $k=1,N$ not given in (\ref{eq:28}).

\subsection{Interpretation as a Majorana fermion code}

It is sometimes instructive to interpret the TFIM code in the language of Majorana fermions.  Defining \begin{subequations}
\begin{align}
    \gamma_i &= Z_1\cdots Z_{i-1}X_i \\
    \xi_i &= Z_1\cdots Z_{i-1}Y_i \; ,
\end{align}
\end{subequations}
where $\gamma$ and $\xi$ are real Majorana modes satisfying $\acomm{\gamma_i}{\gamma_j} = \acomm{\xi_i}{\xi_j} = 2\delta_{ij}$ and $\acomm{\gamma_i}{\xi_j}=0$, we find that
\begin{align}
    U_i = -\frac{i}{\sqrt{2}} \left( \gamma_i\xi_i + \xi_i \gamma_{i+1} \right) \; .
\end{align}
The (inverse) action of a TFIM gate on the Majorana modes is given by
\begin{subequations}
\begin{align}
    U_i \gamma_i U_i^\dagger &= \gamma_{i+1} \\
    U_i \gamma_{i+1} U_i^\dagger &= \gamma_i \\
    U_i \xi_i U_i^\dagger &= -\xi_i \\
    U_i \xi_{i+1} U_i^\dagger &= \xi_{i+1} \; .
\end{align}
\end{subequations}
We see that $\gamma_i \longleftrightarrow \gamma_{i+1}$ while $\xi_i \longrightarrow -\xi_i$ under $U_i$ ($\xi_{i+1}$ is invariant). Due to this linearity, the span of these local Majorana modes will be preserved, so these modes can be a convenient operator basis for analyzing the circuit dynamics. After the encoding procedure $\tilde{U}$, these modes transform as
\begin{subequations}
\begin{align}
    \gamma_i &\longrightarrow \gamma_{\sigma(i)} \\
    \xi_i &\longrightarrow -\xi_i
\end{align}
\end{subequations}
with the exception of $\xi_N$ remaining invariant. The permutation cycle $\sigma$ is given by
\begin{align}\label{eq:cycle}
    \sigma = 
    \begin{cases}
        (1\quad 3\quad 5\quad\dots\quad N-1\quad N\quad N-2\quad N-4\quad\dots\quad 2) & \text{for even }N \\
        (1\quad 3\quad 5\quad\dots\quad N\quad N-1\quad N-3\quad N-5\quad\dots\quad 2) & \text{for odd }N \; .
    \end{cases}
\end{align}
Other than the boundary sites, the cycle $\sigma$ will map odd sites to consecutively increasing odd sites and even sites to consecutively decreasing even sites. Each pair of Majorana modes $\gamma,\xi$ initially start on a single site. The spatial spreading of these modes via $\sigma$ can be interpreted as a measure of entanglement in the system. For multiple applications of $\tilde U$, the dynamics of the Majorana modes can computed via $\mathrm{S}_N$ cycle multiplication rules. For the remainder of the paper we will take $N$ to be even; the case for odd $N$ can be easily inferred by comparing the permutation cycles in the following Eq. (\ref{eq:cycle}).

\section{Quantum error correction for one qubit}\label{sec3}

\begin{figure}[t]
\centering
\includegraphics[width=.75\textwidth]{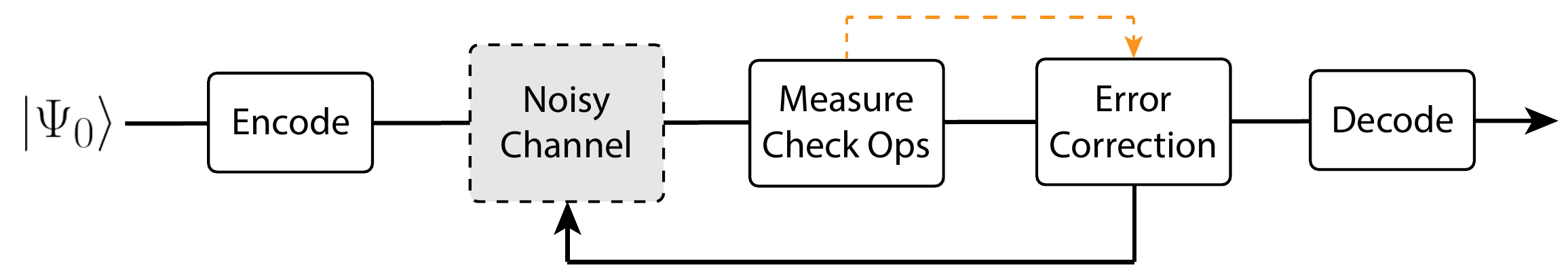}
\caption{The general procedure of stabilizer quantum error correction is depicted in a flowchart. The error correction depends on the error syndrome given by the outcomes of measuring the check operators (orange dashed arrow).}
\label{fig:QEC}
\end{figure}

For this section we will utilize stabilizer error correction to protect one qubit of information. The general idea is that we will measure operators that give us information on undesirable local errors while leaving our quantum state unchanged. This procedure is figuratively illustrated in Fig. \ref{fig:QEC}. Any modifications to the quantum state made by the errors are easy to then manually correct. 

\subsection{Correcting against $Z$ errors a la repetition code}
We will first construct a $\llbracket N,1,1 \rrbracket$ stabilizer quantum error-correcting code, meaning that we encode 1 logical qubit amongst $N$ physical qubits with code distance 1. At time $t=0$, the stabilizer group is generated by $Z_2,\dots,Z_N$.  Any operator which is a product of any of these stabilizers will act trivially on our initial state. The initial logical operators are $X_L = X_1$ and $Z_L = Z_1$. After applying $\tilde{U}$, the stabilizers $Z_2,\ldots, Z_N$ evolve into
\begin{align}\label{eq:check ops}
    \tilde{Z}_k = -i\tilde{U}\gamma_k\xi_k\tilde{U}^\dagger = i\gamma_{\sigma(k)}\xi_k = \begin{cases}
        X_kZ_{k+1}X_{k+2} & \text{for odd $k\neq N-1$} \\
        Y_{k-2}Z_{k-1}Y_k & \text{for even $k\neq 2,N$} \\
        X_{N-1}X_N & \text{for $k=N-1$} \\
        Y_1Y_2 & \text{for $k=2$} \\
        -Y_{N-2}Z_{N-1}Y_N & \text{for $k=N$} \; ,
    \end{cases}
\end{align}
where the tilde denotes the evolved operator via $\tilde{U}$. Similarly, the logical operators become
\begin{subequations}\label{eq:logical ops}
\begin{align}
    \tilde{X}_L &= \tilde{X}_1 = \tilde{U}\gamma_1\tilde{U}^\dagger = \gamma_3 = Z_1Z_2X_3 \\
    \tilde{Z}_L &= \tilde{Z}_1 = -i\tilde{U}\gamma_1\xi_1\tilde{U}^\dagger = i\gamma_3\xi_1 = X_1Z_2X_3 \; .
\end{align}
\end{subequations}
To correct for errors introduced by local projective measurements along the $Z$-axis (alternatively, $Z$ errors), we can use the evolved stabilizer generators in Eq. (\ref{eq:check ops}) as check operators for our encoding. This is quite similar to the conventional repetition code: a local $Z$-measurement and its associated projection operator $P_k^\pm = (1\pm Z_k)/2$ will anticommute with exactly two check operators corresponding to $i$ and $\sigma(i)$. If we organize the sites by their position in the cycle $\sigma$ and measure all check operators simultaneously, we obtain an error syndrome equivalent to that of the quantum repetition code. We then apply the appropriate local phase-flip ($Z$) operators to recover our encoded state.  We can uniquely identify the location of all errors provided there are fewer than $N/2$ of them, and there are no errors in the syndrome measurements themselves. Thus, with high probability (for $N\gg 1$), we can correct for $Z$-measurements as long as the external local measurement probability $p<0.5$ in between the syndrome measurements.  In the presence of faulty syndrome measurements, the success probability of correcting for errors will follow from an identical analysis as the quantum repetition code. Such measurements can arise either due to the environment, or due to the user.  To recover our initial information back onto a single site, we simply apply the inverse encoder $\tilde{U}^\dagger$.

We can in fact protect two qubits of information against $Z$ errors if we wrap our chain in a circle. We take the same configuration of unitary gates as in Fig. \ref{fig:encoder}, but add an additional unitary gate connecting sites $N$ and 1 in the second layer (call this new encoder $\tilde{U}'$). This new encoder corresponds to the permutation cycle
\begin{align}\label{eq:cycle2}
    \sigma' = 
    \begin{cases}
        (1\quad 3\quad 5\quad\dots\quad N-1)\;(N\quad N-2\quad N-4\quad\dots\quad 2) & \text{for even }N \\
        (1\quad 3\quad 5\quad\dots\quad N)\;(N-1\quad N-3\quad N-5\quad\dots\quad 2) & \text{for odd }N \; .
    \end{cases}
\end{align}
Since the new permutation cycle $\sigma'$ is broken into two independent cycles, if we initialize two qubits on an even and odd site separately, then we obtain an $\llbracket N,2,1 \rrbracket$ version of the above stabilizer code with the even and odd sites acting as independent systems. In this circular arrangement, the global $\mathbb{Z}_2$ symmetry \eqref{eq:Zbar} splits into a $\mathbb{Z}_2 \times \mathbb{Z}_2$ symmetry corresponding to the product of $Z$s in the even and odd sectors respectively, in connection with the permutation cycles shown above. Up to local rotations, this encoded state and corresponding error correction is equivalent to that of the 1D cluster state \cite{Briegel_2001} with periodic boundary conditions. If we apply another round of $\tilde{U}'$, the symmetry becomes $\mathbb{Z}_2 \times \mathbb{Z}_2 \times \mathbb{Z}_2 \times \mathbb{Z}_2$ because the permutation $\sigma$ has 4 independent cycles. In general, applying multiple rounds of $\tilde{U}'$ will allow us to encode more logical qubits: one per permutation cycle.

\subsection{Correcting against arbitrary single qubit errors}

The TFIM code can easily be generalized to correct for arbitrary single qubit errors for even $N\ge 10$, resulting in a $\llbracket N\geq 10,1,3 \rrbracket$ stabilizer code.. We begin with the initial state in Eq. (\ref{eq:psi0}) and apply $\tilde{U}'$ twice consecutively to obtain the following check operators:
\begin{align}
    \tilde{Z}'_k = \begin{cases}
        X_kZ_{k+1}Z_{k+2}Z_{k+3}X_{k+4} & \text{for odd $k$} \\
        Y_{k-4}Z_{k-3}Z_{k-2}Z_{k-1}Y_k & \text{for even $k$}
    \end{cases} \; ,
\end{align}
where the site index $k\equiv k\mod N$. The error correction procedure for a local $Z$ error will still be the same as before:  $Z_i$ will anticommute with two check operators. A local $X$ error will in general anticommute with 3 to 5 check operators if it acts on an odd or even site respectively. A lookup table for the types of errors and associated error syndrome is listed in Table \ref{tab:lookup2}. The ability to correct for both $X$ and $Z$ errors can be extended to arbitrary single-qubit errors due to linearity. In order to have the required number of check operators for single-site error correction, we require at least $N=10$ qubits.\footnote{As seen in Table \ref{tab:lookup2}, detecting an error may require spotting a $-1$ syndrome measurement outcome for sites $k$ and $k+4$ (mod $N$).  When $N=8$, we cannot tell whether these sites correspond to $k$ and $k+4$ or $k-4$ and $k$, and thus certain errors cannot be distinguished.  When $N=6$, the $X_{2k}$ and $Y_{2k+1}$ patterns cannot be distinguished, e.g.}  If the check operator length is less than 5, then the corresponding error syndromes will no longer differ from one another by at least 2 checks, and there could be indistinguishable errors depending on the starting position of the logical qubit. According to the quantum Hamming bound, in order to correct for a single arbitrary qubit error with one logical qubit, we need $2^{n-1}-1 \geq 3n$ for $n$ physical qubits. The minimum number of qubits needed is thus 5. Although the TFIM code does not saturate the Hamming bound, it is naturally designed to be implemented in Rydberg atom arrays, as discussed below.

\begin{table}[t]
\centering
\begin{tabular}{|c|cccccc|}
\hline
\textbf{Local Error} & \multicolumn{6}{c|}{\textbf{Check operators}} \\ \hline\hline
 & $\tilde{Z}'_{2k-3}$ & $\tilde{Z}'_{2k-1}$ & $\tilde{Z}'_{2k+1}$ & $\tilde{Z}'_{2k}$ & $\tilde{Z}'_{2k+2}$ & $\tilde{Z}'_{2k+4}$ \\ \hline
$X_{2k}$ & 1 & 1 & 0 & 1 & 1 & 1 \\
$X_{2k+1}$ & 0 & 1 & 0 & 0 & 1 & 1 \\
$Y_{2k}$ & 1 & 1 & 0 & 0 & 1 & 0 \\
$Y_{2k+1}$ & 1 & 1 & 1 & 0 & 1 & 1 \\
$Z_{2k}$ & 0 & 0 & 0 & 1 & 0 & 1 \\
$Z_{2k+1}$ & 1 & 0 & 1 & 0 & 0 & 0 \\ \hline
\end{tabular}
\caption{Types of local errors and their associated error syndromes are shown after two rounds of $\tilde{U}'$. An entry of 1 indicates that the error anticommutes with the above check operator and 0 if it commutes.  Note that each row above differs from another row in $\ge 2$ columns; thus, even if one of the $Z^\prime$ in the table (which would be known to the user) represents a logical $Z$ (and should not be measured), it will still be uniquely possible to identify an error given any pattern of syndrome measurements.  }
\label{tab:lookup2}
\end{table}

For this code to protect against arbitrary X or Z errors, it was crucial to have the $N$ and 1 qubits interacting.  To understand why, let us suppose the logical qubit is 1, and we apply the unitary $\tilde U^k$ (i.e. $k$ rounds of $\tilde U$).  The check operators are $\gamma_{\sigma^k(i)}\xi_i$ for $i>1$, and we detect errors by deducing the number of these checks which do not commute with the error.  There is only one check operator which has a Pauli on site 1:  $\tilde Z_{2k} = Y_1 Z_2 \cdots  Y_{2k} $, since $\tilde Z_1 = X_1Z_2\cdots X_{2k+1}$ is logical and cannot be measured.  We deduce, therefore, that if we measured the check operators and found that all returned $+1$ except for $\tilde Z_{2k}\rightarrow -1$, we would not be able to tell apart an $X_1$ or $Z_1$ error; thus we could not correct the error.   Alternatively, if we measure $Y_1$, we will destroy the logical qubit, since no logical $X$ operator ($\tilde X_1$, $\tilde X_1 \tilde Z_{2k}$, etc.) can commute with $Y_1$.  In the circular arrangement, this problem is solved because there are multiple check operators which have $Z_1$s in their Pauli string.

Another perspective is that the code distance is bounded by the minimum logical operator length for local errors. Without periodic boundary conditions, $Y_1$ will remain invariant, and so the code distance is 1 after encoding. This issue is solved with periodic boundary conditions since now we have gates which can evolve $Y_1$ into longer Pauli strings.

\subsection{Quantum teleportation}
\begin{figure}
\centering
\includegraphics[width=.35\textwidth]{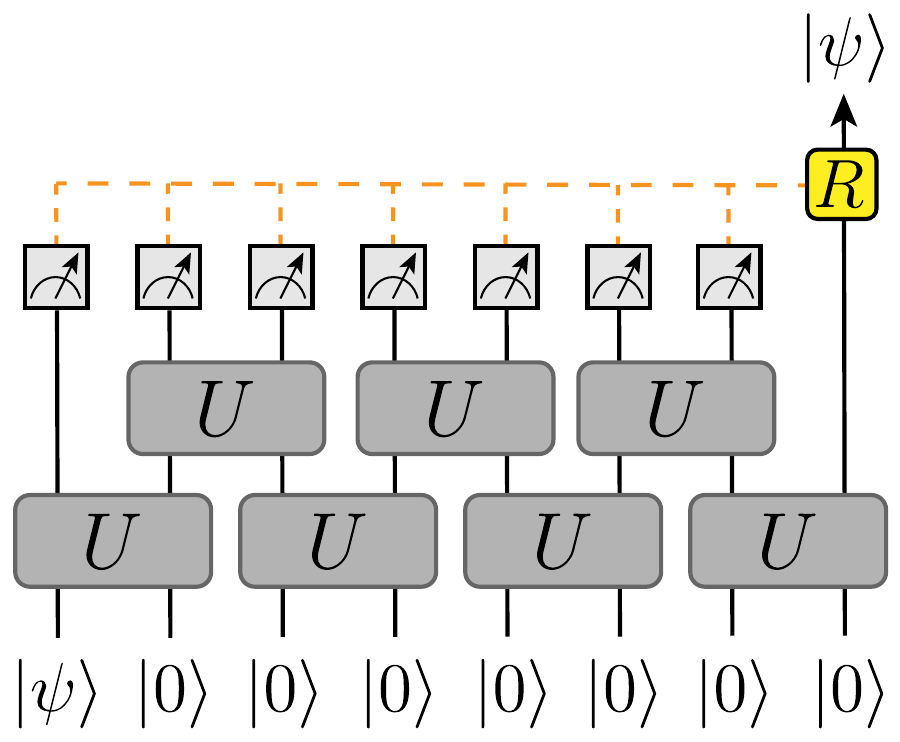}
\caption{Illustration of the state transfer protocol for $N=8$ sites. The initial information is stored on the leftmost site. The orange dashed line denotes the classical information channel required to perform the correct local rotation gate $R$ at the end. The arrowed boxes represent local projective measurements along the $z$-axis.}
\label{fig:teleport}
\end{figure}

Just as the TFIM code can be used to correct for errors, it can also be used to perform a measurement-assisted quantum teleportation between any two qubits in the system.   Suppose we wish to transfer our initial quantum state, located on the left end of the 1D chain ($i=1$), to the rightmost qubit of our 1D chain $(i=N)$. By using local projective measurements and classical communication, we can achieve such state transfer after a single layer of the TFIM code.  Indeed, examining Eq. (\ref{eq:PC state}) suggests that measuring $N-1$ sites will push the information onto the last site up to a local $X$ or $Z$ rotation (depending on the measurement outcomes). The state transfer protocol is illustrated in Fig. \ref{fig:teleport}. 

We now explain how to correct for both $X$ and $Z$ error in this process.  To correct for a possible bit flip ($X$) error, we note that the $Z$-projection operators commute with the global symmetry operator $\bar{Z}$. Thus, we simply need to compute the overall parity after the measurements. If the parity of the $N-1$ measurement outcomes is even (even number of down spins measured), no application of $X$ is required on the final site. If that parity is odd, then we need to apply an $X$ to correct for the resulting bit flip.

In addition to the bit flip error, we could also have a phase flip ($Z$) error on our final state.  Recall that such a phase flip error was not possible in the repetition code; the possibility of this phase flip can be understood as the ``price" to pay for the ability to correct for both X and Z errors (as in Section \ref{sec3}), as well as for the ability to generate the state after a finite-depth circuit (we will elaborate on this in a future paper).  We can determine whether a phase flip has occurred by manipulating our check and logical operators in Eqs. (\ref{eq:check ops}) and (\ref{eq:logical ops}) respectively. Starting from our initial logical operators, we can create longer (Pauli) strings of logical operators by successively multiplying check operators. The goal is to create a new logical operator which commutes with all $N-1$ projection operators and thus remains a good logical operator. For the initial $X_1$ logical operator, this problem can be interpreted as trying to ``move" the $X$ operator onto the final site.  As an example, if we start on site 1 and wish to transfer our information onto site 8 after one application of $\tilde{U}$, we can define a new initial logical operator $X_L = X_1Z_3Z_5Z_7$ such that
\begin{align}\label{eq:new logical}
    \tilde{X}_L = \underbrace{Z_1Z_2X_3}_\text{logical} \underbrace{(X_3Z_4X_5)(X_5Z_6X_7)(X_7X_8)}_\text{checks} = Z_1Z_2Z_4Z_6X_8 \; .
\end{align}
The value of the string of Pauli $Z$'s on the left of the $X$ will take on $\pm 1$. If $Z_1Z_2Z_4Z_6$ has odd parity (-1), then we need to apply a $Z$ gate on site 8. For the logical $Z_L$, we can multiply by all stabilizer generators $Z_2,\dots,Z_8$ to obtain $\bar{Z}$, which remains invariant under the dynamics. Thus, measuring sites 1 to 7 in this example and applying a final local gate based on parities of measurement outcomes will allow us to achieve state transfer:
\begin{subequations}\begin{align}
    \tilde{X}_L &\xrightarrow{\text{ LOCC }} X_8 \\
    \tilde{Z}_L &\xrightarrow{\text{ LOCC }} Z_8 \; ,
\end{align}\end{subequations}
where LOCC is short for local operations and classical communication.

In the Majorana representation, we are multiplying check operators in such a way that we obtain pairs of Majoranas $\gamma_k\xi_k$ on sites other than the final site: in the example above, \begin{equation}
    \underbrace{\gamma_3}_\text{logical}\underbrace{(\xi_3\gamma_5)(\xi_5\gamma_7)(\xi_7 \gamma_8)}_\text{checks} 
\propto Z_1Z_2Z_4Z_6X_8
\end{equation}When applying the unitary $\tilde U$, we observe from the structure of the Majorana modes that we could \emph{always} arrange to have either, neither, or both of $\gamma$ and $\xi$ on every single site except for one, by suitable multiplication of the check operators $\gamma_{\sigma(i)}\xi_i$, simply because we choose the checks to multiply by in order to ``follow" the cycle from the logical operator $\gamma_i$ to the final site of interest ($\gamma_j$).  The number of check operators we must multiply by is $\ell$, where $\sigma^\ell(i)=j$.  The Majorana representation makes clear that with sufficient knowledge, information is only truly lost in this code if \emph{every single site is measured}.  This is true for every value of $N$, since the cycle in (\ref{eq:cycle}) connects every single site.

With the above scheme, we are able to protect a bit of quantum information for long times as well as achieve rapid state transfer to extract the information onto a single site. Note that the decoding step at the end is not bound by the Lieb-Robinson theorem \cite{Lieb1972} (information travels at a finite velocity in a local quantum spin chain under unitary dynamics such as $\tilde U$) since it involves non-unitary projective measurements across the entire chain.  One major advantage of our code over the standard repetition code is that the Lieb-Robinson Theorem proves the unitary dynamics necessary to generate the state (\ref{eq:QRC}) scales as $t\sim N$.\footnote{This can be seen by noting that (\ref{eq:QRC}) is a GHZ-like state in the $X$-basis, and the time to prepare such states is constrained by Lieb-Robinson bounds.  \cite{tran2021} contains some discussions on the relationship between GHZ prepration time and Lieb-Robinson bounds.}  Our code not only is capable of achieving what the quantum repetition code can, but it can be implemented much faster, and (as we showed above) can even correct for non-commuting errors (both $Z$ and $X$).

\section{Higher distance codes}\label{sec5}
Thus far, we have shown how the TFIM code can protect against arbitrary single qubit errors.  We now argue that with some modifications, TFIM-based codes can also protect a finite fraction of qubits against a finite fraction of stochastic erasure errors, and that they may represent a more easily implementable version of the random low-depth codes analyzed in \cite{gullans} with comparable performance.

In an operator language, if we want qubits on sites $k_1,\ldots, k_n$ to be protected, then we must be able to find check operators such that $(\tilde{X},\tilde{Z})_{k_j}\times \prod \tilde{Z}$ is the identity on sites with errors for suitable products of check operators, and each of the logical $X$ and $Z$ on the protected sites.  

Unfortunately, as given, the TFIM code cannot protect two logical qubits against arbitrary errors.  To understand why, observe that on any given site, there are at most 2 \emph{logical or check} operators that can have an X on that site.  If at any point in time during the code, each of these operators (e.g. $X_1Z_2X_3$ and $X_3Z_4X_5$) happen to both be logical operators, then measurement or erasure on site 3 will destroy some information: only the combined logical operator $(X_1Z_2X_3)(X_3Z_4X_5)$ can survive erasure (by being proportional to identity $I$ on site 3): all other check operators will have Pauli Z on site 3.  This conclusion extends to codes generated by higher depth circuits (i.e. applying $\tilde U^k$ for $k>1$), for the same reason (albeit the notation gets more cumbersome).  If we wish to protect a finite \emph{fraction} of qubits, we will inevitably run into a pair of qubits which cannot simultaneously be protected.  Hence, we will look for a modified TFIM code in which the check operators have a reasonable density of both Pauli $X$s and $Z$s to overcome this obstruction.

\begin{figure}
\centering
\includegraphics[width=.4\textwidth]{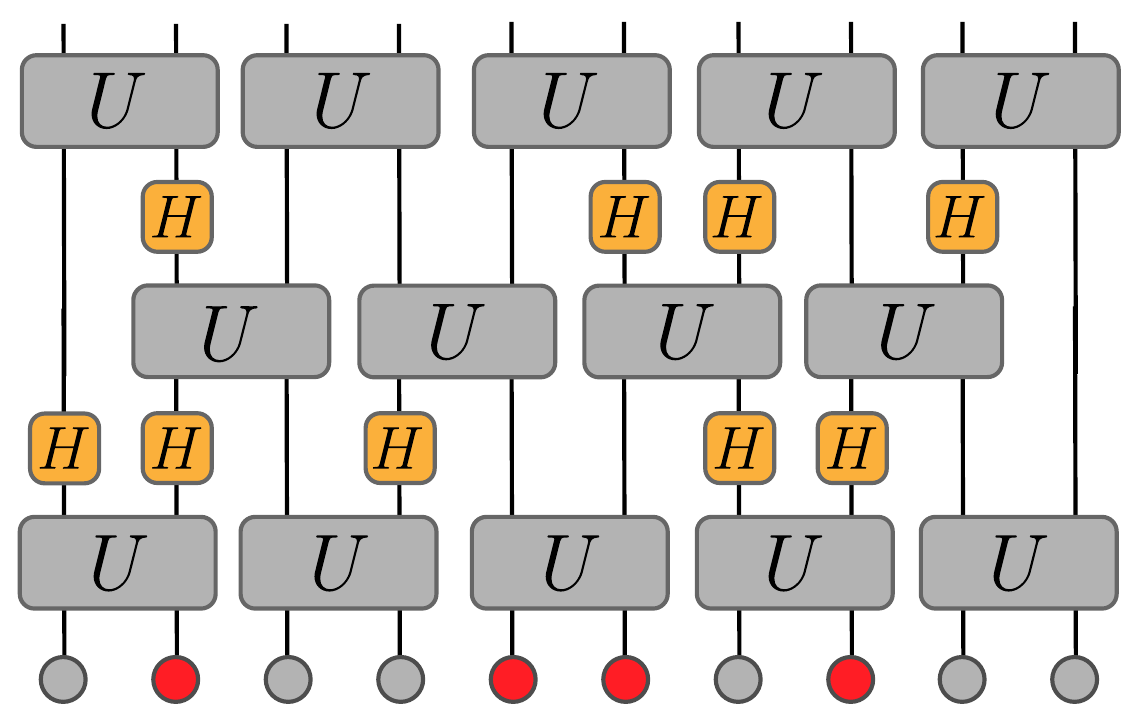}
\caption{A modified TFIM encoder is illustrated with intermediate layers of random local Hadamard ($H$) gates. The red colored sites denote initial logical qubits, and the light-gray colored sites are initialized in the $\ket{0}$ state.}
\label{fig:randH}
\end{figure}

Fortunately, there is a relatively simple strategy to address this, which is to apply Hadamard gates periodically to each qubit. The simplest implementation is to apply Hadamard gates with local probability $p_H$ in an intermediate circuit layer between the encoding layers, depicted in Fig. {\ref{fig:randH}}. The logical and check operators will then be more complicated strings of $X$ and $Z$.  As our circuits will remain Clifford circuits under these local Hadamard gates (which simply flip $X$ and $Z$ on the site they are applied),  we can readily numerically simulate the distribution of Paulis in our check operators.

\begin{figure}[t]
\centering
\includegraphics[width=.32\textwidth]{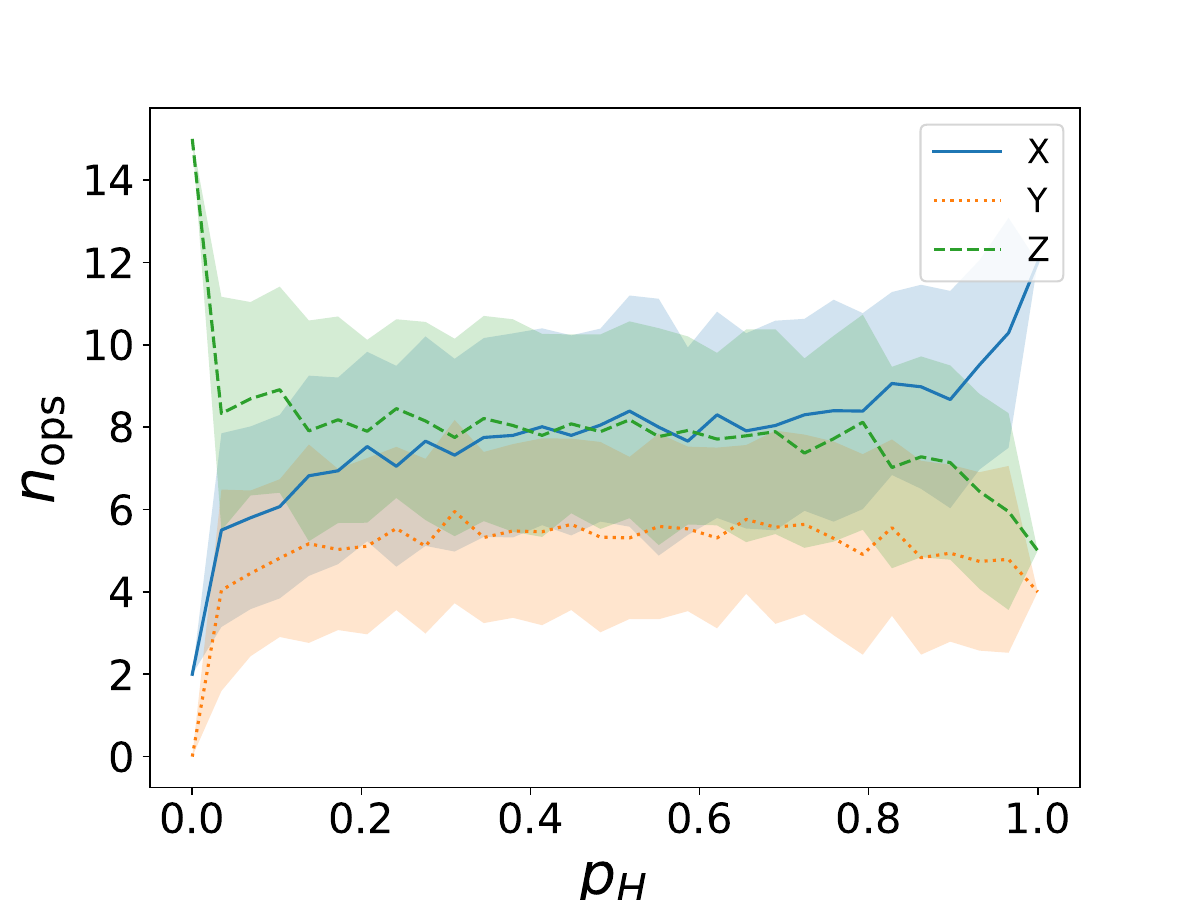}
\includegraphics[width=.32\textwidth]{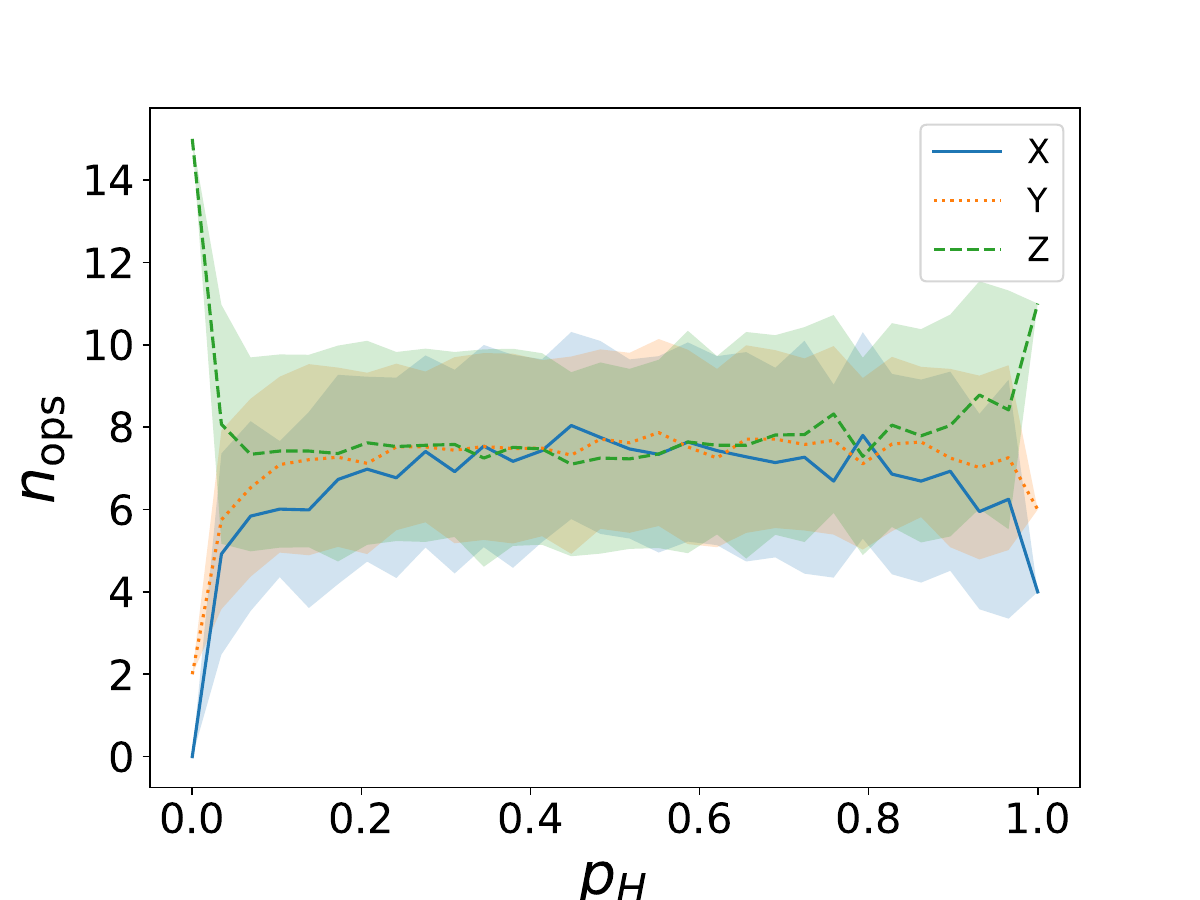}
\includegraphics[width=.32\textwidth]{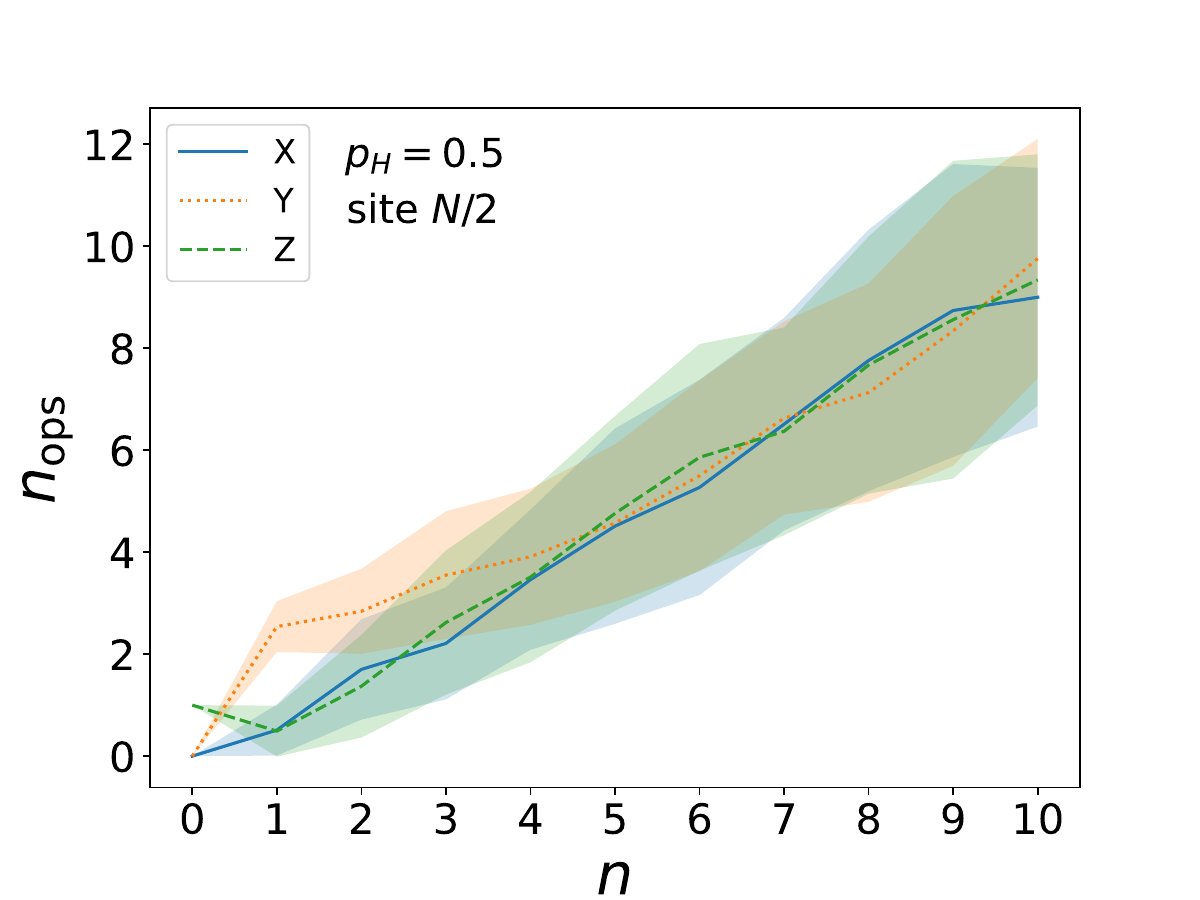}
\caption{The average number of check operators with $X,Y,Z$ support on an even (left plot) or odd site (middle plot) as a function of local Hadamard probability $p_H$ is shown for $n=4$ rounds of $\tilde{U}$ and $N=100$. The average number of overlapping check operators on the central site of the chain is also plotted as a function of $n$ for $p_H=0.5$ (right plot). The averages are performed over $100$ circuit iterations, and error bars (denoted with shading) represent the standard deviation of sample-to-sample fluctuations.}
\label{fig:XZ ops}
\end{figure}

To analyze the performance of this modified TFIM code, we consider a prototypical error model of random erasure errors.  Namely, on each site, with some probability $p_E$, we measure a random Pauli matrix and discard the measurement outcome.  We assume that we have knowledge of which sites are erased in this way. In order to protect against such a measurement, we need to make sure that all logical operators can be chosen to be the identity on every single erased site.  We find that in typical realizations of such a code, there are many $\tilde Z_i$ operators that can be used to correct for arbitrary single-site errors (see Fig. \ref{fig:XZ ops}), implying that some fraction of them could correspond to logical qubits.  We do not present an explicit analysis of the fraction of qubits which can be protected given the probability a given site is erased.  However, as suggested in Figure \ref{fig:XZ ops}, after $n$ rounds of this random TFIM code, for most values of $p_H$ there are at least $\mathrm{O}(n)$ check operators that could be used to remove an $X,Y$ or $Z$ on any given site.

If the erasure errors are random, we need to worry that there will be a rare sequence of many sites in a row in which all sites in the sequence are erased.  If the erasure probability is $q$, then the longest sequence of erased sites will have a length $L$ of order $q^L \sim N^{-1}$ (recall that the total number of qubits is $N$); namely, \begin{equation}
    L \sim \frac{\log N}{\log q^{-1}}. \label{eq:logN}
\end{equation} 
This suggests that if we want to use a finite fraction of qubits in the system as logical, we will need to run the code for a time $t\propto L \propto \log N$ to ensure that we can correct against all erasure errors in the system, if the probability that a given site is erased is sufficiently low. In other words, we need the lengths of our check operators to be at least $\mathrm{O}(L)$ for sufficient erasure protection. Previous studies of 1D codes indeed show that check operators must be extensively large to correct for an extensive amount of errors \cite{Haah_2012}.

\begin{figure}[t]
\centering
\includegraphics[width=.4\textwidth]{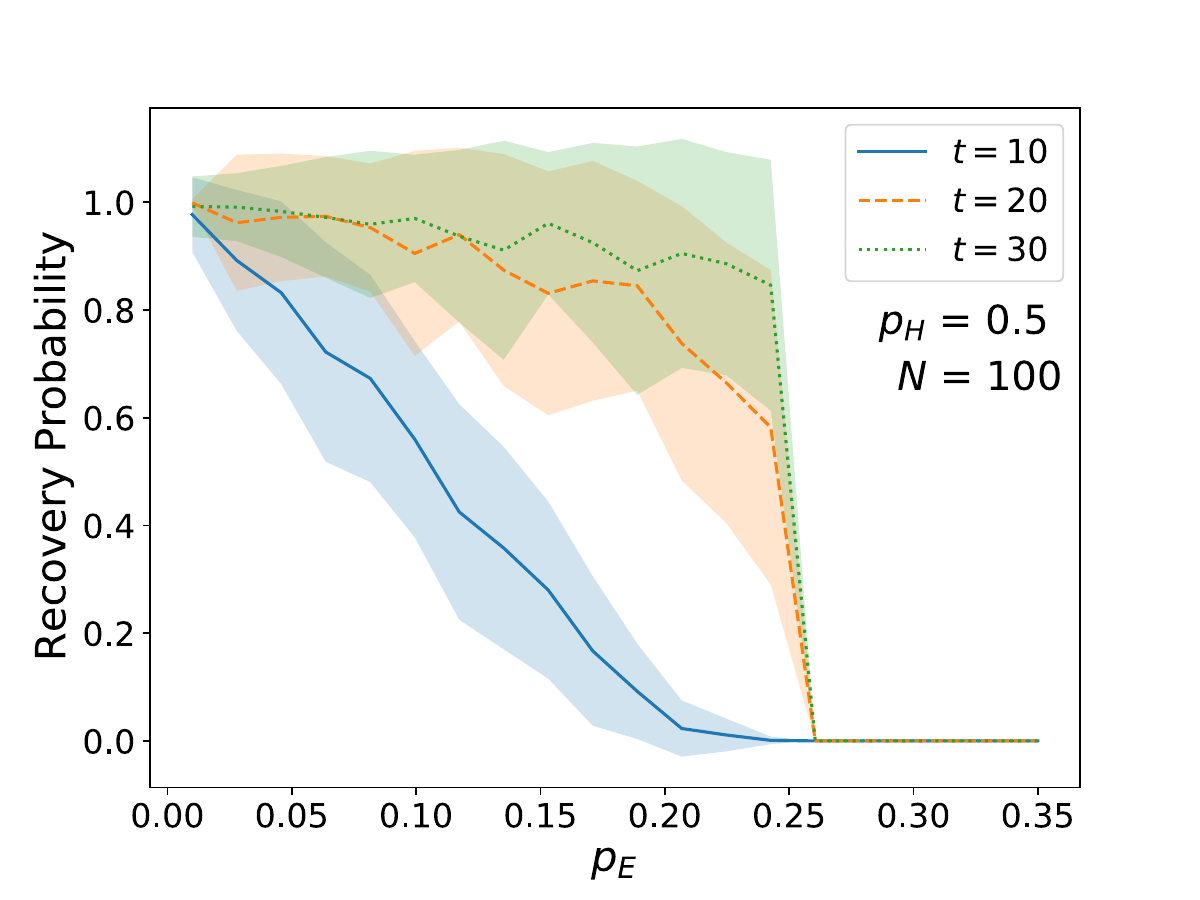}
\includegraphics[width=.4\textwidth]{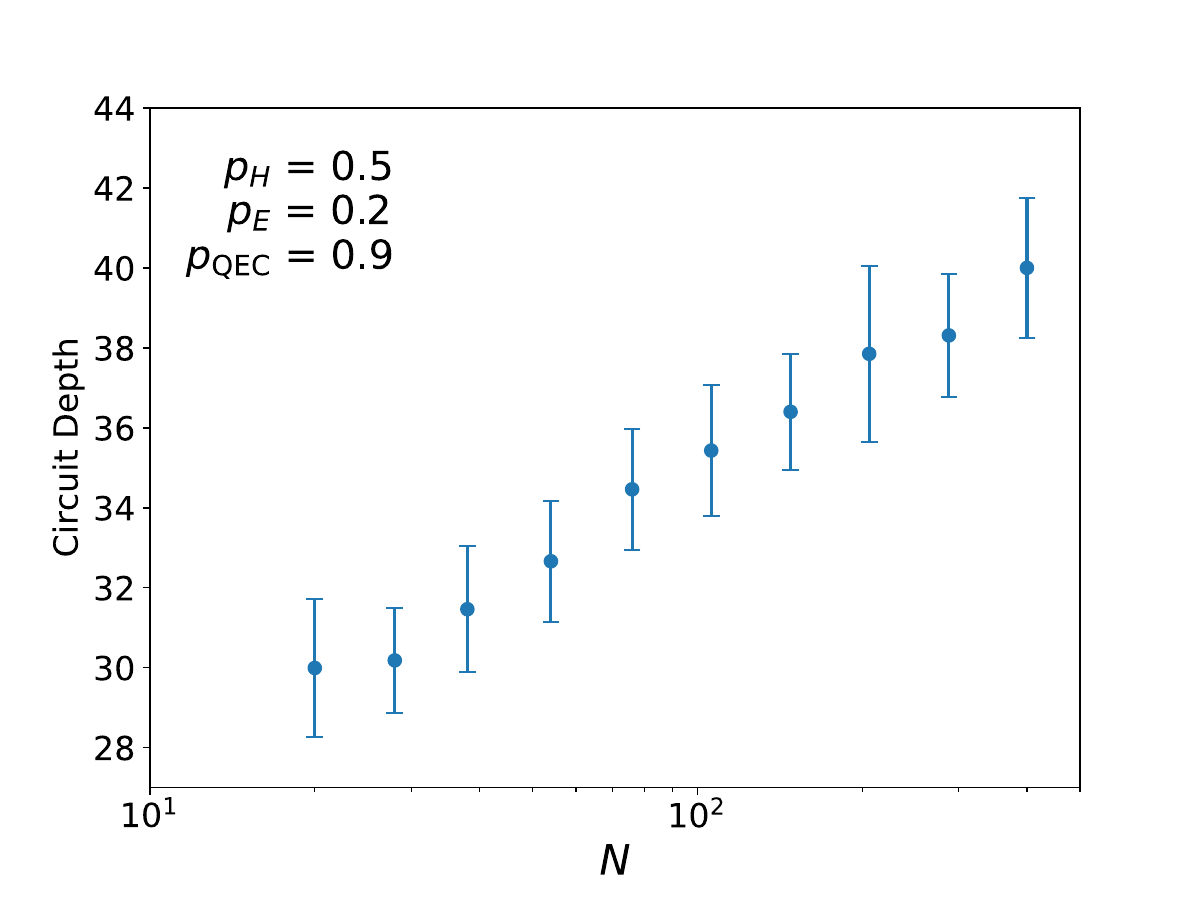}
\caption{Numerical simulation results are shown for a TFIM code with random Hadamard gates and erasures. $N/2$ qubits are logical, and so only $N/2$ check operators are available for erasure protection. Left: Successful erasure recovery probability is plotted as a function of single site erasure probability for three different circuit depths $t=2n$ for $n$ rounds of $\tilde{U}$. Note the quantum Hamming bound sets an upper limit $p_E \leq (1-f)/2$ for protecting $fN$ logical operators; for this simulation we have $f=1/2$. Right: The circuit depth $t$ required to recover a given erasure probability $p_E=0.5$ with average success probability $\expval{p_\mathrm{QEC}}=0.9$ is plotted as a function of the length of the chain (note the log scale on the x axis). Both plots are obtained by averaging over 100 iterations of the circuit.}
\label{fig:depth_vs_N}
\end{figure}

A more detailed statistical analysis of a similar problem was presented in \cite{gullans} in the context of random Clifford circuits, which are believed to be the (asymptotically) lowest depth one-dimensional circuits capable of error correction.  Our numerics suggest that the TFIM code can also saturate this asymptotic bound, while arguably having a more obvious experimental implementation.  While we leave a detailed statistical analysis of the performance of our code in the $N\rightarrow \infty$ limit to future work,  Fig. \ref{fig:depth_vs_N} shows that our code can protect against a finite density of erasure errors on $N$ sites after a circuit depth which scales as $\log N$, consistent with the heuristic argument in (\ref{eq:logN}).

\section{Implementation in Rydberg atom arrays}\label{sec6}
\begin{figure}
    \centering
    \includegraphics[scale=.5]{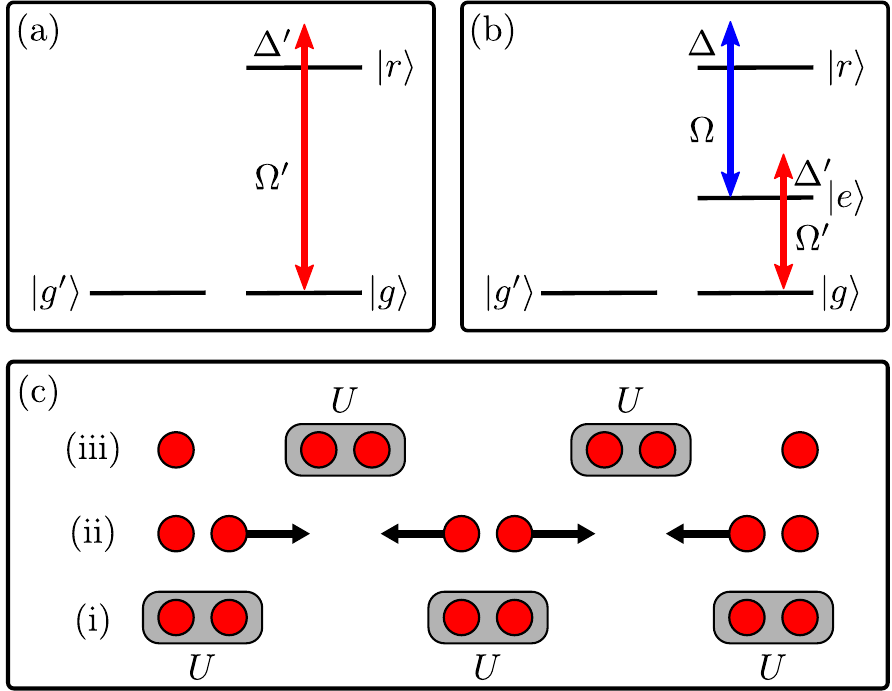}
    \caption{Approaches for engineering the encoding gates using Rydberg interactions. The states $|g\rangle, |r\rangle , |e\rangle$ denote a ground state, Rydberg state, and an intermediate state, respectively. The long-lived state $|g'\rangle$ is used to store the qubit (along with $|g\rangle$) when the gates are not being applied. The drive $\Omega'$ with detuning $\Delta'$ is used to both generate the transverse field and to remove a longitudinal field, which arises due to the form of the Rydberg interactions. In both, we define $|0\rangle \equiv |g\rangle$ and store $|1\rangle$ in the long-lived state $|g'\rangle$ when the gates $U$ are not being applied, coherently transferring $|g'\rangle$ to/from the interacting $|1\rangle$ state via a fast $\pi$ pulse to apply $U$. (a) In the first approach, we define $ |1\rangle \equiv |r\rangle$, leading to vdW Ising interactions. 
    (b) In the second approach, the $|e\rangle$ and $|r\rangle$ states are weakly dressed with Rabi frequency $\Omega$ and detuning $\Delta \gg \Omega$, producing a dressed state $|d\rangle \approx |e\rangle + \frac{\Omega}{2 \Delta} |r\rangle$. We define $|0\rangle \equiv |d\rangle$, leading to soft-core Ising interactions with a vdW tail. (c) In order to avoid long-range interactions between different pairs of atoms, the atoms are moved in (ii) between the first (i) and second (iii) layers of the encoding, eliminating unwanted interactions.}
    \label{fig:rydberg}
\end{figure}

We now describe the realization of the TFIM quantum error correcting code using Rydberg atoms. To do so, we shall consider two different approaches, using $1/r^6$ van der Waals (vdW) interactions: (1) using the Rydberg interactions directly and (2) using dressed Rydberg interactions \cite{Pupillo2010,Johnson2010,Henkel2010, Jau2016,Zeiher2016,Zeiher2017, Arias2019, Borish2020,Guardado-Sanchez2021, Schine2021}, both of which are illustrated in Fig.~\ref{fig:rydberg}. In the first approach, the Rydberg interactions can be very strong, allowing for quick implementation of the necessary gates. However, because of the $1/r^6$ scaling of the vdW interactions, this requires knowing the distance of the atoms with a high degree of accuracy. In the second approach, the interaction potential plateaus at short distances, avoiding this issue. However, the gate times are increased, so decoherence from dissipation becomes more relevant, particularly from avalanche processes due to blackbody radiation \cite{Goldschmidt2016,Aman2016}.
In both approaches, cross-talk between parallel two-qubit gates can be eliminated by moving the atoms [cf.~Fig.\ref{fig:rydberg}(c)], which can be achieved using Rydberg tweezer arrays \cite{Bluvstein2021}.

In the first approach [Fig.~\ref{fig:rydberg}(a)], the qubit is encoded via $|0 \rangle \equiv |g\rangle$ and $|1 \rangle \equiv |r\rangle$, where $|g\rangle, |r\rangle$ denote ground and Rydberg states, respectively.  The vdW interactions from the Rydberg interactions take the form
\begin{equation}
\begin{aligned}
    V_{\text{vdW}} &= \sum_{i < j} \frac{C_6}{r_{ij}^6} |r_i r_j \rangle \langle r_i r_j| \\
    &= \frac{1}{4}\sum_{i < j} \frac{C_6}{r_{ij}^6} (Z_i Z_j + Z_i + Z_j + 1), \label{eq:Vvdw}
\end{aligned}
\end{equation}
where $C_6$ denotes the strength of the vdW interactions, $r_{ij}$ is the distance between atoms $i$ and $j$. Note that we have assumed no angular dependence in $C_6$, which can be achieved by either using a $L=0$ state or by fixing $\theta_{ij} = \theta$, which is defined relative to the quantization axis. Before and after the gate is applied, the $|1\rangle$ state can be coherently transferred from/to some other long-lived state $|g'\rangle$ via a fast $\pi$ pulse to suppress decoherence.

In the second approach [Fig.~\ref{fig:rydberg}(b)], we utilize Rydberg dressing by introducing a third atomic state $|e\rangle$ which is weakly dressed with $|r\rangle$ via a drive with Rabi frequency $\Omega$ and detuning $\Delta \gg \Omega$. Here, we now encode $|1\rangle \equiv |d\rangle \approx |e\rangle + \frac{\Omega}{2 \Delta} |r\rangle$, where $|d\rangle$ is one of the two dressed states of the drive. As a result of the dressing, the Rydberg interactions take the modified form
\begin{equation}
\begin{aligned}
    V_{\text{vdW}} &= \frac{\Omega^4}{8 \Delta^3} \sum_{i < j} \frac{1}{1+ (r/r_b)^6} |d_i d_j \rangle \langle d_i d_j| \\
    &= \frac{\Omega^4}{32 \Delta^3} \sum_{i < j} \frac{1}{1+ (r/r_b)^6} (Z_i Z_j + Z_i + Z_j + 1),
\end{aligned}
\end{equation}
where $C_6/r_b^6 = - 2 \Delta$ defines the blockade radius $r_b$. Due to the dressing, the interactions take the form of a soft-core potential with a power-law tail. As a result, the interaction is approximately constant for a range of $r \lesssim r_b$. Like in the first approach, we assume the there is no angular dependence in $C_6$. As in the first approach, the $|1\rangle$ state can be mapped from/to a long-lived state $|g'\rangle$ to suppress decoherence.

Using either of the above approaches, the Ising interactions may be prepared. To realize the desired Hamiltonian, we must add a transverse field and remove the longitudinal field $Z$, both of which can be achieved using a drive applied to the first atom and by going to a rotating frame for the second atom. For the first approach, the $|g\rangle \to |r \rangle$ transition is driven, while for the second approach, the $|g\rangle \to |e\rangle$ transition is driven. Note that for the second approach, light shifts due to the weak dressing field should be taken into account to turn the $|g\rangle \to |e\rangle$ drive into a resonant $|g\rangle \to |d\rangle$ drive. In both cases, this introduces a $\Omega' X/2$ term for the driven atom, providing the desired transverse field. 
To remove the longitudinal field, we apply a detuning $\Delta'$ to the drive which exactly cancels the longitudinal field on the driven atom. For the undriven atom, we may simply use the same rotating frame defined by $\mathcal{R}_2 = \mathrm{e}^{-\mathrm{i}\Delta' Z_2 t/2}$ as for the driven atom. Since this commutes with the Hamiltonian, it only removes the longitudinal field.  Hence
\begin{subequations}
    \begin{equation}
        H = V_\text{vdW} + H_{\mathrm{d}},
    \end{equation}
    \begin{equation}
        H_{\mathrm{d}} = \frac{\Omega'}{2} X_1 - \frac{\Delta'}{2} Z_1,
    \end{equation}
    \begin{equation}
        \tilde{H} = \mathcal{R}_2^\dagger H \mathcal{R}_2 - \mathrm{i} \mathcal{R}_2^\dagger \partial_t \mathcal{R}_2 = V_0 Z_1 Z_2 + \left(V_0 - \frac{\Delta'}{2}\right) Z_1 + \left(V_0 - \frac{\Delta'}{2}\right) Z_2 + \frac{\Omega'}{2} X_1,
    \end{equation}
\end{subequations}
where $H_\mathrm{d}$ is the drive used to generate the transverse field (already in the corresponding rotating frame), $\tilde{H}$ is the Hamiltonian in the rotating frame, and $V_0$ is the Rydberg interaction between the two atoms. From this, we see that by setting $\Omega = \Delta = 2 V_0$, we may realize the desired Hamiltonian.

In either of the above approaches, we can suppress the dephasing effects of the power-law tails in interactions in an optical tweezer set-up by simply moving the atoms quite close together before applying the $U_i$ gates.  In general, we might expect that in a time $t$, there is possible dephasing by an angle (assuming the vdW interactions are given by (\ref{eq:Vvdw}), though an analogous formula holds for the other case): \begin{equation}
    \phi \propto 4 t \sum_{n=1}^\infty \frac{C_6}{(n r_1)^6} < 4.08 \frac{C_6}{r_0^6} t \propto 4.08 \left(\frac{r_0}{r_1}\right)^6.
\end{equation}
Here $r_0$ ($r_1$) is the typical inter-atom spacing between neighboring atoms which are (not) subject to $U$ [cf.~Fig.~\ref{fig:rydberg}(c)], and we have used the fact that the gate time $t$ is inversely proportional to $C_6/r_0^6$. By increasing the ratio $r_1/r_0$, this effect can be arbitrarily decreased. For example, using $r_1/r_0 = 4$, the effect is suppressed by at least a factor of 4000 compared to $r_1/r_0 = 1$.

In addition to engineering $U$ to encode the qubit, we must also measure the check operators in order to implement the error correcting code. This may be achieved through the use of an ancilla qubit for each check operator \cite{Auger2017,Bluvstein2021,Cong2022,Wu2022}. For each qubit in a given check operator, we must then apply a two-qubit entangling gate with the ancilla gate, along with any necessary one-qubit gates. Although the check operators involve several qubits, we may again take advantage of the ability to move the ancilla qubits as needed. The requisite two-qubit entangling gates can be naturally realized via Rydberg blockade gates \cite{Jaksch2000,Saffman2005,Saffman2010,Xia2013}, which have been demonstrated experimentally with high fidelities \cite{Levine2018,Levine2019,Graham2019,Madjarov2020,Zhang2020}. In particular, these allow for a straightforward implementation of a controlled-NOT (CNOT) gate. 

As an example, if we wish to measure the check operator $X_k Z_{k+1} Z_{k+2} Z_{k+3} X_{k+4}$, we may proceed as follows: (1) Initialize the ancilla qubit in $|0\rangle$ and apply a $\pi/2$ pulse to the $k$ and $k+4$ qubits, mapping $X$ to $Z$. (2) Apply a Rydberg CNOT gate between the ancilla qubit and each qubit in the check operator. (3) Apply a $-\pi/2$ pulse to the $k$ and $k+4$ qubits, mapping $Z$ back to $X$. (4) Measure the ancilla qubit; the measurement result corresponds to the parity of the check operator. Note that the CNOT gates for each ancilla qubit may be applied in parallel as long as one ensures that cross-talk between the different gates is eliminated, as discussed above for the implementation of $U$.

Since the check operator length is larger than the code distance, there is a possibility of cascading errors during the syndrome extraction. In order to combat this effect and achieve fault tolerance, we would need to utilize a more complicated scheme such as Shor's cat-state syndrome extraction \cite{Shor_1996}.

\section{Conclusions}
In this paper we have introduced a simple quantum error correcting code based on a time-dependent transverse field Ising model.  This represents a practical alternative and improvement to the quantum repetition code, realizable in near term platforms, including trapped Rydberg atoms, where we proposed concrete implementations of the code.   

This code is a simple illustration of more profound and general concepts which have been recently discussed in the literature, such as the ability to perform quantum teleportation across arbitrarily large distances using only measurements and a finite depth circuit acting on an initial product state \cite{Bao:2021con}, and the link between SPT phases and error correction \cite{2004.07243,Verresen:2021wdv,Tantivasadakarn:2021vel}.  A careful analysis of the dynamics in this model has also suggested general trade-offs between the sensitivity of the code to measurement outcomes and the time required to generate it: general theorems along these lines have recently been reported \cite{speed_limit_meas}.

\section*{Acknowledgements}
We thank Emanuel Knill for extremely valuable discussions, and Victor Albert for pointing out \cite{Pastawski:2016ggn}.  This work was supported by a Research Fellowship from the Alfred P. Sloan Foundation under Grant FG-2020-13795 (AL), by the U.S. Air Force Office of Scientific Research under Grant FA9550-21-1-0195 (YH, AL), by the NIST NRC Research Postdoctoral Associateship Award (JTY), and NIST (AMK).

\bibliography{thebib}

\end{document}